\documentclass{aa}  
\let\linenumbers\relax
\usepackage{graphicx}
\usepackage{hyperref}
\usepackage{txfonts}
\usepackage{amsmath}

\graphicspath{{plots/}}

\newcommand{\hGpc}{h^{-1}~{\rm Gpc}}

\newcommand{\hMpc}{{\ifmmode{,h^{-1}{\rm Mpc}}\else{$h^{-1}$Mpc}\fi}}
\newcommand{\hkpc}{{\ifmmode{,h^{-1}{\rm kpc}}\else{$h^{-1}$kpc}\fi}}
\newcommand{\hMsun}{{\ifmmode{\,h^{-1}{\rm {M_{\odot}}}}\else{$h^{-1}{\rm{M_{\odot}}}$}\fi}}

\newcommand{\Mstar}{{\ifmmode{,M_{*}}\else{$M_{*}$}\fi}}
\newcommand{\Mhalo}{{\ifmmode{\,M_{\rm halo}}\else{$M_{\rm halo}$}\fi}}

\newcommand{\simba}{\textsc{Gizmo-Simba}}
\newcommand{\theth}{\textsc{The Three Hundred}}
\newcommand{\ahf}{\textsc{AHF}}

\newcommand{\gadgetx}{\textsc{Gadget-X}}

\defcitealias{Cui2018}{C18}

\begin{document}
    \title{\theth: Gas Properties Outside of Galaxy Cluster with the WHIM Contribution and Detection}
    
    \author{
        Renjie Li \inst{1,2,3} 
        \and 
        Weiguang Cui \inst{3,4,5} 
        \and 
        Ang Liu \inst{6,7} 
        \and 
        Huiyuan Wang \inst{1,2} 
        \and 
        Atulit Srivastava \inst{3,4} 
        \and 
        Romeel Dave \inst{5,8}
        \and
        Frazer R. Pearce \inst{9}
    }
    
    \institute{
            Key Laboratory for Research in Galaxies and Cosmology, Department of Astronomy, University of Science and Technology of China, Hefei, Anhui 230026, China\\
            \email{phylrj@mail.ustc.edu.cn}
        \and
            School of Astronomy and Space Science, University of Science and Technology of China, Hefei 230026, China
        \and
            Departamento de Física Teórica, M-8, Universidad Autónoma de Madrid, Cantoblanco 28049, Madrid, Spain
        \and
            Centro de Investigaci\'{o}n Avanzada en F\'{i}sica Fundamental (CIAFF), Facultad de Ciencias, Universidad Aut\'{o}noma de Madrid, E-28049, Madrid, Spain
        \and
            Institute for Astronomy, University of Edinburgh, Royal Observatory, Edinburgh EH9 3HJ, United Kingdom
        \and
            Institute for Frontiers in Astronomy and Astrophysics, Beijing Normal University, Beijing 102206, China
        \and
            Max Planck Institute for Extraterrestrial Physics, Giessenbachstrasse 1, 85748 Garching, Germany
        \and
            Department of Physics and Astronomy, University of the Western Cape, Robert Sobukwe Rd, Cape Town 7535, South Africa
        \and
            School of Physics and Astronomy, University of Nottingham, Nottingham NG7 2RD, UK
    }

    \abstract
    {We investigate the physical properties and detectability of warm-hot intergalactic medium (WHIM) gas with temperatures in the range $10^5<T<10^7$~K around galaxy clusters using simulated galaxy clusters from \theth\ project. In simulations with different input physics (\simba\ and \gadgetx), we consistently find that the median gas temperature decreases to the WHIM upper bound, $10^7$~K, at $\sim 2 \times R_{200c}$, while the WHIM mass fraction increases with radius until $\sim 3\times R_{200c}$, where it plateaus at $\sim 70$ per cent. By simulating X-ray emission from all gas components, we find that the WHIM contribution to the soft X-ray band (0.2–2.3~keV) increases with radius but eventually plateaus at larger distances. The differences between the two simulations become more pronounced at higher redshifts and larger radii. Finally, after accounting for observational effects, primarily by removing (sub)halos, we predict that the signal-to-noise ratio of the X-ray signal obtained by stacking the eRASS1 galaxy cluster catalogue will be $\sim 7$ for \simba\ and $\sim 21$ for \gadgetx.}


   \maketitle

\section{Introduction} \label{sec:intro}

Measurements of the cosmic microwave background (CMB) indicate that baryons comprise approximately $\Omega_b \sim 0.049$ of the cosmic mass-energy budget \citep{1Planck2020A&A}. Estimates from the Lyman-$\alpha$ forest at $z \sim 2-3$ support this value \citep{Rauch1998ARA&A}. However, summing all observed baryonic components across different wavelengths in the local Universe reveals that $\sim 30-50\%$ of these baryons remain undetected, leading to the so-called `missing baryon problem' \citep[e.g.,][and references therein]{2Cen1999ApJ, 3Bregman2007ARA&A, 4Shull2012ApJ}. This discrepancy may arise from incomplete accounting of all baryons. For instance, stellar mass is quantified through optical surveys, hot gas is detected via X-ray and thermal Sunyaev-Zel'dovich (tSZ) measurements \citep{17Sunyaev1972CoASP}, while radio observations trace cold gas \citep[e.g.,][]{Battye2004MNRAS}.

Cosmological hydrodynamical simulations have long suggested that the missing baryons reside in the warm-hot intergalactic medium (WHIM), a diffuse gas component outside galaxy halos with temperatures of $T\sim 10^5-10^7$~K, which is extremely challenging to detect \citep[see e.g.,][]{12Dave2001ApJ, 13Martizzi2019MNRAS, 14Holt2022MNRAS, 15Li2022ApJ, 16Gouin2023A&A}. Additionally, the bulk of the WHIM is thought to reside within cosmic filaments, prompting focused observational efforts \citep[e.g.,][]{Cui2019}.

Detecting and quantifying the WHIM is difficult because it contains little atomic hydrogen observable in \ion{H}{1} absorption and is highly diffuse, leading to an extremely low surface brightness in emission. Furthermore, most absorption and emission features are expected in the soft X-ray band, where instrumental sensitivity is low and foreground contamination is significant.

Various methods have been attempted to detect the WHIM. Using quasar absorption spectra from XMM-Newton, \cite{Nicastro2018} reported WHIM detection via \ion{O}{7} and \ion{O}{8} absorption in a few sightlines. However, this required significant telescope time and cannot be scaled up to obtain robust statistics until next-generation facilities such as NewAthena \citep{22Stofanova2024MNRAS} become available. Meanwhile, \ion{O}{6} probes slightly cooler gas, $T\sim 10^5$~K \citep[e.g.,][]{Tepper2011,Oppenheimer2016,Bradley2022}, and is predominantly found outside virialized regions such as galaxies, groups, or clusters \citep[e.g.,][]{Cen2001,Appleby2023}. Thus, detections of \ion{O}{6} absorbers \citep[e.g.,][]{Tripp2008,Pachat2016MNRAS,Ahoranta2020A&A,Ahoranta2021A&A,HolguinLuna2024} may also indicate the presence of WHIM gas. Stacking techniques \citep{23Kovacs2019ApJ} have also been employed, but they make it difficult to constrain the physical conditions of the absorbers, limiting confidence in their association with the WHIM.

The WHIM can also be probed via its thermal SZ decrement. Simulations suggest that most warm-hot gas resides in cosmic filaments \citep[e.g.,][]{9Cui2018MNRAS, 6Tuominen2021A&A, 10Galarraga2021A&A, 11Galarraga2022A&A}. Accordingly, \cite{20deGraaff2019A&A} stacked Planck tSZ Compton $y$-parameter maps for over one million galaxy pairs to detect the WHIM in intergalactic filaments, obtaining a mass fraction of $\sim 11\%$. Similar results were found using slightly different methods and datasets \citep{18Tanimura2019MNRAS, 19Tanimura2020A&A, Isopi2024}. \cite{5Ferraro2016PhRvD} explored the use of kinetic SZ signals as an unbiased tracer of the large-scale electron distribution. However, since the SZ signal strength depends on gas pressure, which declines rapidly outside halos (albeit not as steeply as X-ray emission), detecting individual WHIM systems remains challenging until next-generation SZ facilities come online. 

Another approach involves stacking soft X-ray emission from putative WHIM gas, but this too is difficult due to the WHIM's diffuse nature and moderate ionization \citep{24Ursino2011MNRAS, 25Parimbelli2023MNRAS}. Recently, \cite{Zhang2024A&A} reported a $5.4\sigma$ detection of the WHIM in filaments.

WHIM gas is also expected in the outskirts of galaxy clusters, where the gas density and temperature naturally fall within the WHIM range as they decrease with radius \citep[see, e.g.,][]{LiQY2023}. \cite{26Bonamente2022MNRAS} analyzed ROSAT data of the Coma cluster, identifying an excess of soft X-ray radiation consistent with WHIM gas. eROSITA has also detected gas near the Coma cluster and surrounding filaments \citep{27Churazov2021A&A, 29Churazov2023MNRAS, 28Reiprich2021A&A}, while \cite{McCall2024A&A} used eROSITA to trace gas in the Virgo cluster out to $3R_{200}$. However, these detections typically reach a maximum statistical significance of only $\sim 3\sigma$. Moreover, contamination from bright sources and background fluctuations complicates the analysis of low-redshift systems, making stacking necessary for studying cluster outskirts at higher redshifts.

\cite{30Zhang2024arXiv} analyzed hot circumgalactic medium (CGM) profiles of low-mass halos out to $R_{\mathrm{vir}}$, finding a lower-than-expected hot CGM fraction within $R_{\mathrm{vir}}$, which reapproaches the $\Lambda$CDM prediction near $3R_{\mathrm{vir}}$. In one of the most comprehensive recent studies, \cite{Lyskova2023MNRAS} stacked 38 galaxy clusters at $0.05 < z < 0.2$ with masses $2\times10^{14}M_{\odot} < M_{500c} < 9\times10^{14}M_{\odot}$, obtaining X-ray profiles extending to $3R_{500c} \sim 1.9R_{200c}$. They evaluated density, temperature, and entropy profiles, and estimated a baryon fraction of $\Omega_b/\Omega_m \approx 0.15$ outside $R_{500c}$. These results suggest that X-ray stacking around clusters may currently be the most viable approach for detecting WHIM emission with existing instruments.

In our previous work \citep{15Li2022ApJ}, we used constrained realizations to generate 2D X-ray maps and profiles of a mock Coma cluster. The predicted emission line distributions were consistent with cluster observations \citep{31Mirakhor2020MNRAS}, supporting the presence of WHIM gas in cluster outskirts. Here, we extend this work by analyzing \theth\ simulations \citep{32Cui2018MNRAS}, which contain 324 clusters, to develop methods for detecting WHIM using current instruments via stacking. Since most simulation models calibrate parameters against observed data from within clusters, their accuracy in diffuse regions remains uncertain. Therefore, we compare results from two different galaxy formation models, \simba\ and \gadgetx, in the cluster outskirts.

\autoref{sec:data} describes our simulation data, halo catalogue, and X-ray modelling. The density and temperature properties of the clusters are analysed in \autoref{sec:theo}. \autoref{sec:obs} presents mock observations using a stacking technique to study the X-ray evolution of warm-hot gas beyond clusters. In particular, \autoref{subsec:obs_detect} details our methodology for estimating signal-to-noise ratios from observational data. Finally, we discuss our findings and conclusions in \autoref{sec:con}.

\section{Data and Method} \label{sec:data}

We introduce \theth\ simulations in \autoref{subsec:simu} and describe the halo catalogue constructed using Amiga's Halo Finder (\ahf) \citep{Knollmann2009} in \autoref{subsec:halo}. The X-ray emission from simulated gas is computed using our C++ program, which interpolates tables from AtomDB \citep{Foster2020}. The code is designed for computational efficiency while maintaining accuracy within computational constraints. Our methodology is detailed in \autoref{subsec:xray}.

\subsection{The Three Hundred Project} \label{subsec:simu}
\theth\ project selects 324 distinct regions centered on the most massive clusters from the dark-matter-only MultiDark Planck 2 simulation (MDPL2; \cite{Klypin2016}). MDPL2 consists of $3840^3$ dark matter particles within a $(1~\hGpc)^3$ volume and adopts the Planck 2015 cosmology \citep{PlanckCollaboration2016}. These clusters were subsequently re-simulated using several full-physics hydrodynamical models, starting from modified initial conditions at $z=120$. Gas particles were introduced in a central high-resolution region, while beyond 15~\hMpc, the particle resolution was reduced. The high-resolution particles have masses of $M_{dm} = 1.27 \times 10^9~\hMsun$ for dark matter and $M_{gas} = 2.36 \times 10^8~\hMsun$ for gas.

In this work, we use simulation results from \gadgetx\ \citep{Rasia2015, Beck2016}, a modified version of GADGET3 (an updated version of GADGET2; \cite{Springel2005}), and \simba\ \citep{Dave2019, Cui2022}. Further details on the simulations can be found in \citep[e.g.,][]{32Cui2018MNRAS, Cui2022}.

\subsection{Halo catalogue and sample selection}
\label{subsec:halo}
Halos in the simulation are identified using Amiga's Halo Finder (\ahf; \cite{Knollmann2009}) with an overdensity threshold of $200\rho_{\text{crit}}$. The selected cluster samples from MDPL2 remain centered in high-resolution regions at $z=0$ by construction and are thus re-matched with the central halos. At redshift zero, the central halos are defined as the most massive ones within a search radius of approximately $R_{200c}$ from the center of the high-resolution region. At higher redshifts, the central halos are traced through the merger tree to identify their main progenitors.

For observational comparisons, 2D maps and profiles are generated. In each simulation, the analysis is conducted within a circular region centered on the central halo. This cylindrical volume has a radius of $5R_{200c}$ from the halo center and a half-thickness of $5R_{200c}$. Since \theth{} simulations are zoom-in simulations, only the selected high-resolution regions are used in this work to exclude low-resolution particles.

\subsection{X-ray map generation and stacking procedure} \label{subsec:xray}
The X-ray surface brightness is computed using a C++ program, updated from the version used in our previous work \citep{15Li2022ApJ}. The simulation region is divided into small pixels with a solid angle $\Delta\Omega$, and the surface brightness is calculated as:
\begin{align}
\begin{split}
S_x(E_1, E_2)
&=\frac{1}{\Delta\Omega}\sum_i\frac{1}{4\pi D^2_L(z_i)} 
\\
&\int_{E_1(1+z_i)}^{E_2(1+z_i)}\Lambda(E,T_i,Z_i)n_{{\rm H},i}N_{{\rm e},i}dE,
\end{split}\label{eq_xray}
\end{align}
where $E_1$ and $E_2$ define the X-ray photon energy range. The emissivity, $\Lambda$, is obtained from look-up tables provided by AtomDB \citep[][version 3.0.9]{Foster2020}\footnote{http://www.atomdb.org/index.php}. The variables $n_{{\rm H},i}$, $N_{{\rm e},i}$, $T_i$, and $Z_i$ represent the number density of hydrogen ions, the number of free electrons, the gas temperature, and the metallicity of the gas element $i$, respectively. The terms $z_i$ and $D_L(z_i)$ denote the cosmological redshift and luminosity distance of the gas element. We adopt the $0.2 \sim 2.3$~keV energy band to maximize the signal from warm-hot gas while matching eROSITA's sensitivity range. The code is optimized for large-scale simulations where spectral resolution is not required, enabling efficient computation of X-ray maps for thousands of clusters within hours. Key optimizations include precomputing spectral interpolation and performing band integration during preprocessing rather than at runtime, significantly improving computational efficiency. However, observational effects and systematics are not yet included. When stacking X-ray maps, it is sometimes necessary to remove halos to isolate the properties of the warm-hot gas. To achieve this, we excise all pixels containing \ahf-identified halo gas, which introduces holes in single-cluster maps. During stacking, these empty pixels are excluded, meaning holes are zero-filled when computing fluxes. Radial profiles are computed using the median of individual cluster measurements to ensure robustness. 

\section{Physical properties of gas in the vicinity of central clusters} \label{sec:theo}
\subsection{Temperature and density} \label{subsec:Trho}

\begin{figure*}
    \centering
    \includegraphics[scale=0.55]{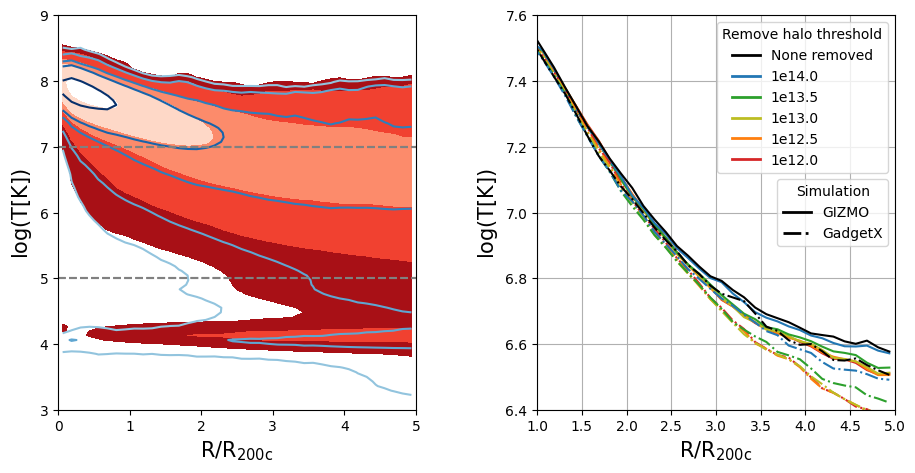}
    \caption{Temperature distribution of \simba\ (color-filled contours) and \gadgetx\ (contour lines). The left panel shows the temperature distribution of all gas as a function of normalized radius from the cluster centers, while the right panel presents the corresponding temperature profiles. In the left panel, contour lines enclose 20\%, 50\%, 80\%, 95\%, and 99\% of gas particles within the analysis region, with 2D radii normalized to $R_{200c}$. Horizontal dashed lines indicate the warm-hot temperature range, defined between $10^5$~K and $10^7$~K. The right panel displays temperature profiles for different halo mass removal thresholds as indicated.
    \label{fig:Tprofile}}
\end{figure*}

As shown in \autoref{eq_xray}, temperature and density are key parameters in X-ray calculations. Temperature affects the gas ionization state, the spectral distribution of emitted photons, and the WHIM mass fraction. The left panel of \autoref{fig:Tprofile} presents the temperature distributions as a function of 2D radius for \simba\ (colour-filled contours) and \gadgetx\ (contours).

For this analysis, each simulation is divided into radial and logarithmic temperature bins, counting gas mass in each bin. The results are then stacked across all 324 clusters, with all halos retained in the left panel but removed as indicated in the right panel. Contours represent cumulative gas particle fractions at levels of 20\%, 50\%, 80\%, 95\%, and 99\%.

The median gas temperature decreases from $\sim 10^8$~K at the cluster center to $\sim 4 \times 10^6$~K at $5R_{200c}$. Within $2.5R_{200c}$, \gadgetx\ and \simba\ show only minor temperature differences. However, at larger radii \simba\ forms slightly hotter gas compared to \gadgetx.

The right panel of \autoref{fig:Tprofile} presents the temperature profiles of \simba\ and \gadgetx\ as a function of 2D cluster-centric distance. Mass-weighted temperature profiles are computed before taking the median across all clusters. Data for $R < R_{200c}$ is excluded (see \cite{LiQY2023,LiQY2020} for details on intra-cluster gas properties).

The various profiles illustrate how the gas properties change when halos above different mass thresholds are removed. However, differences between these threshold variants remain minimal out to $R \sim 2.5R_{200c}$. Diffuse gas in \gadgetx\ appears slightly cooler than in \simba, consistent with the contours shown in the left panel. In both simulations, gas transitions into the WHIM phase beyond $\sim 2.2R_{200c}$ and maintains WHIM-level temperatures ($T > 10^5$~K) out to $5R_{200c}$. These results confirm that most gas remains in the WHIM phase even at large radii, with the contribution from halo gas, across a range of mass thresholds, having only a minor impact on the median temperature.

\begin{figure*}
    \centering
    \includegraphics[scale=0.6]{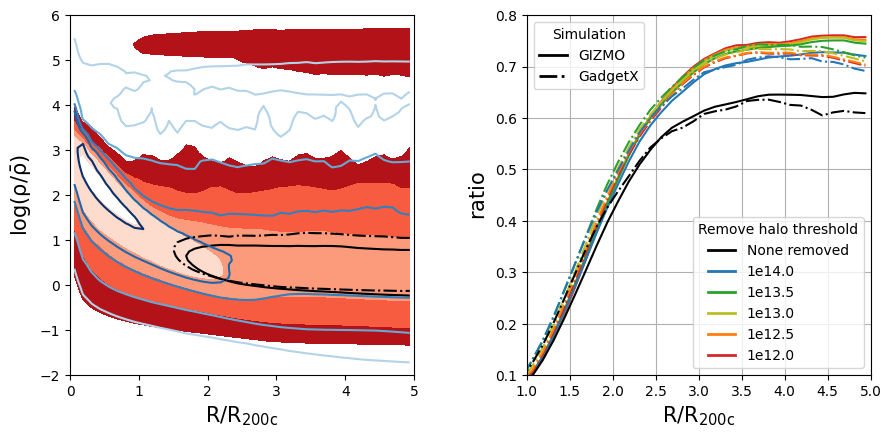}
    \caption{
    Left panel; similar to \autoref{fig:Tprofile}, but for the density distribution of all the gas particles for \simba\ (colour-filled contours) and \gadgetx\ (contours). The contour lines enclose 20\%, 50\%, 80\%, 95\%, and 99\% of gas particles within the analysis region, with 2D radii normalized to $R_{200c}$. 
    The two black contours indicate the region on this plane that encloses 50\% of the WHIM gas in each simulation.
    Right panel: the radial profile of the fraction of gas that is WHIM, coloured lines show different halo mass removal thresholds as indicated in the legend.
    \label{fig:rhocontour}}
\end{figure*}

Gas density, which primarily determines the strength of X-ray emission, is shown as a function of radius in the left panel of \autoref{fig:rhocontour}. Following the methodology of \autoref{fig:Tprofile}, we analyze gas particles in radial shells and stack the results. The mean density is defined as:
\begin{equation} \bar{\rho} = \frac{3H_0^2}{8\pi G} \Omega_{m,0} \label{rhom}\,. \end{equation}
High-density gas beyond $R_{200c}$, typically associated with $T\sim10^4$~K gas, is mostly found in galaxies or other virialized structures. This gas is denser and slightly hotter in \simba\ compared to \gadgetx, where it does not exhibit a clear separation from lower-density gas. Overall, the gas density declines steeply within $2R_{200c}$ and then flattens.

At $R\gtrsim 2R_{200c}$, \gadgetx\ contains a higher fraction of dense gas than \simba, as indicated by the offsets in the 80\%, 95\%, and 99\% contours. This may result in stronger X-ray signals in \gadgetx\ compared to \simba. Regarding the WHIM, this gas in \gadgetx\ also appears slightly denser than in \simba, as highlighted by the two black contour lines.

The right panel of \autoref{fig:rhocontour} shows the WHIM mass fraction as a function of radius, with the y-axis representing the ratio of WHIM gas to total gas mass. This fraction is computed in 2D radial bins across all 324 clusters before stacking. Black lines include all gas, while coloured lines exclude particles within halos above certain mass thresholds. The clear separation between the black and coloured lines, around $\sim 10\%$ at $R \gtrsim 3R_{200c}$, is due to the decrease in total gas mass compared to WHIM mass, as WHIM gas is largely absent inside massive halos (\autoref{fig:Tprofile}). Removing halo gas increases the WHIM fraction by $\sim 10\%$, suggesting that the X-ray signal is more strongly influenced by warm gas rather than hotter gas components.

Declining temperatures (\autoref{fig:Tprofile}, right panel) correlate with increasing WHIM fractions up to approximately $3R_{200c}$ (\autoref{fig:rhocontour}), beyond which WHIM gas comprises $\sim 70\%$ of the total mass. The \gadgetx\ model exhibits a slightly higher WHIM fraction than \simba\ within $2.5R_{200c}$, but this trend reverses at larger radii. This difference may arise from differences in feedback models; for instance, the jet-driven feedback in \simba\ can expel gas to larger radii, reaching distances as far as $4R_{\text{vir}}$, as noted by \cite{Yang2023}. This is broadly consistent with the higher gas temperatures observed in \simba\ (\autoref{fig:Tprofile}, left panel).
Both simulations show maximum WHIM fractions when removing halos below $10^{13.5} M_{\odot}/h$, indicating that WHIM gas constitutes a significant fraction of the total gas mass in lower-mass halos.

\subsection{The redshift evolution of gas properties in galaxy cluster outskirts} \label{subsec:redshift}

\begin{figure}
    \centering
    \includegraphics[scale=0.55]{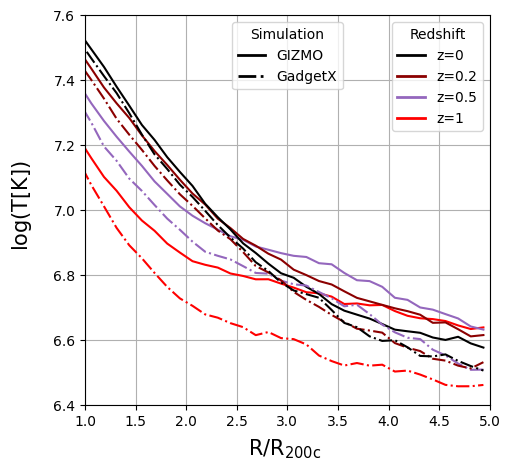}
    \caption{Similar to the right panel of \autoref{fig:Tprofile}, but showing temperature profiles at different redshifts with all halos included. Line styles indicate the simulation type, while colours represent different redshifts. All the profiles show the median temperature of all 324 clusters. In general, gas outside clusters is cooler in \gadgetx\ than in \simba, particularly at large radii.
    \label{fig:Tprofile_redshift}}
\end{figure}

As structure forms, the gas temperature systematically increases within $2.5R_{200c}$, as shown in \autoref{fig:Tprofile_redshift}. This is primarily driven by halo virialization, which appears to have a consistent effect beyond $R_{200c}$. Beyond $2.5R_{200c}$, \simba\ exhibits higher temperatures at $z=0.5$ than at $z=0$, a trend also observed in \gadgetx\ beyond $3R_{200c}$. This difference is partly influenced by the radial normalization. The temperature discrepancies between \simba\ and \gadgetx\ become more pronounced at higher redshifts. For example, at $z=1$, gas temperatures beyond $2R_{200c}$ in \simba\ are nearly twice as high as those in \gadgetx. These temperature differences may be attributed to the strong AGN jet feedback in \simba, which is typically active around $z \sim 2$ \citep{Sorini2022MNRAS}.

\begin{figure}
    \centering
    \includegraphics[scale=0.4]{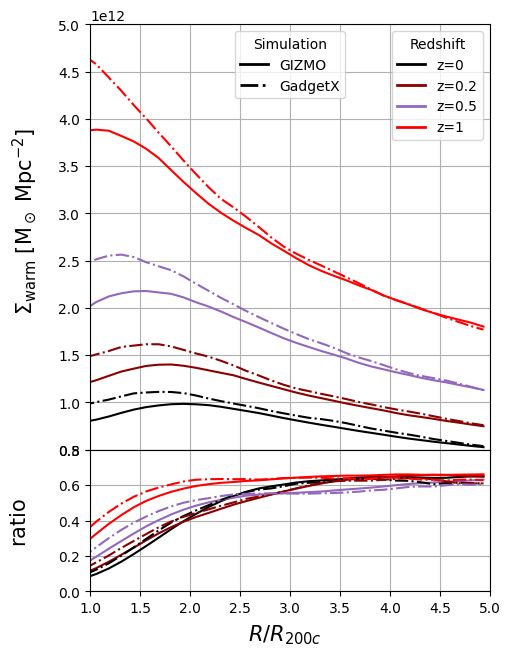}
    \caption{Radial surface density profiles of warm gas (top) and the fraction of warm gas relative to all gas (bottom). Line styles indicate the simulation type, while colours represent different redshifts. Note that no halos have been removed.
    \label{fig:rhoprofile_redshift}}
\end{figure}

Similarly, WHIM gas density profiles at $z=0, 0.2, 0.5, 1$ are shown in \autoref{fig:rhoprofile_redshift}. The top panel indicates higher gas densities at higher redshifts, consistent with cosmic expansion. WHIM gas density is systematically higher in \gadgetx\ than in \simba, with the difference decreasing at larger radii. This may be due to the fact that no halos have been removed in these profiles. As shown in \autoref{fig:rhocontour}, \gadgetx\ exhibits a higher WHIM contribution from low-mass halos compared to \simba.

The bottom panel reveals that WHIM fractions increase with redshift within $\sim 2R_{200c}$. This redshift evolution is primarily driven by the steady increase in gas temperature, as shown in \autoref{fig:Tprofile_redshift}. Beyond $R \gtrsim 2.5R_{200c}$, the WHIM fraction remains remarkably stable, indicating a co-evolution of WHIM and the total gas component that is largely unchanged across redshifts. This stability suggests that stacking cluster outskirts at different redshifts is a reliable approach for studying the WHIM.

\section{X-ray results} \label{sec:obs}
While simulations indicate significant WHIM contributions in cluster outskirts, observational detection remains challenging and requires stacking techniques to improve the signal-to-noise ratio \citep{Veronica2024arxiv, Popesso2024arxiv}. In this section, we assess the feasibility of detecting WHIM gas beyond galaxy clusters using X-ray observations.

\subsection{X-ray map stacking and WHIM detection in the vicinity of galaxy clusters} \label{subsec:obs_overview}
\begin{figure*}
    \centering
    \includegraphics[scale=0.55]{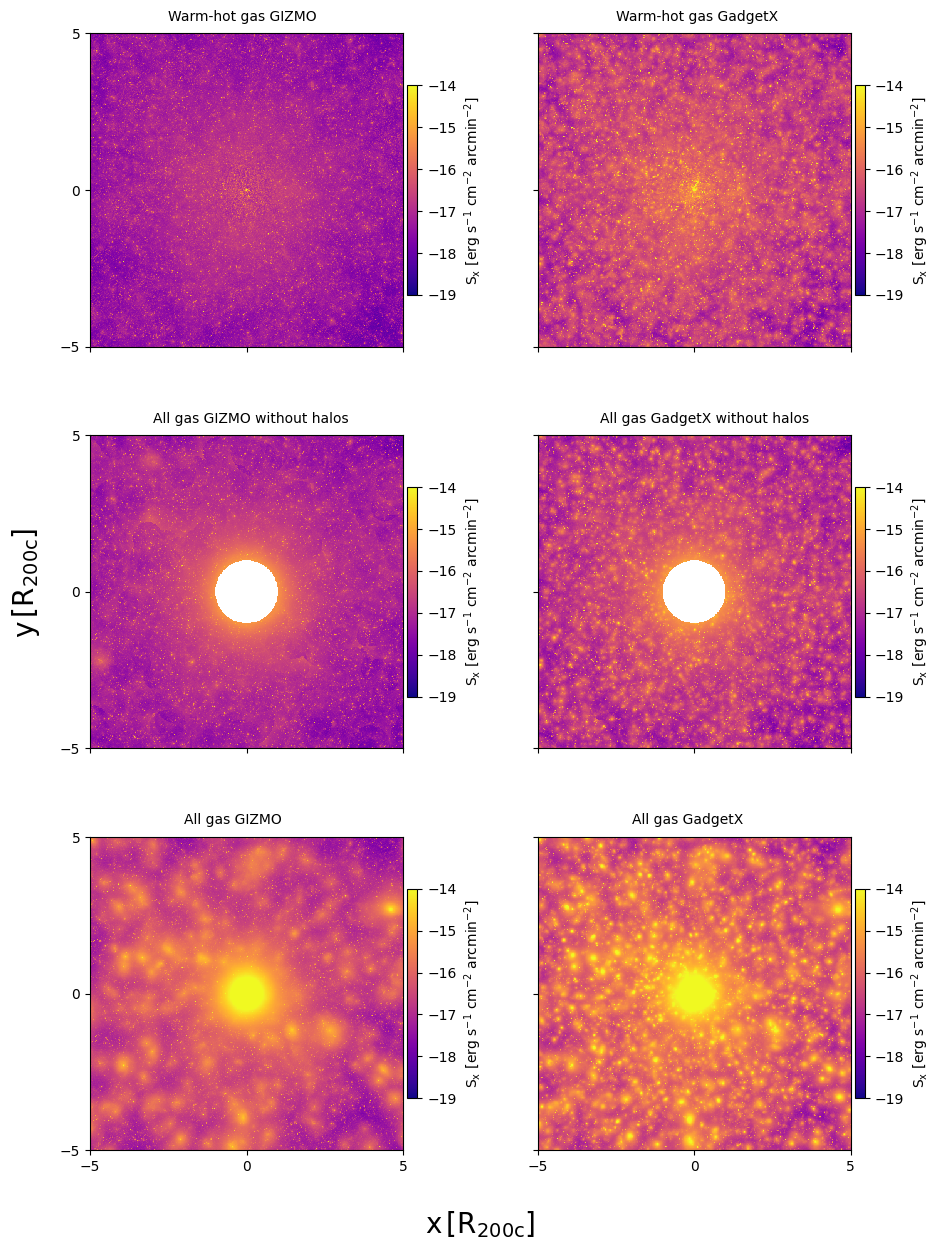}
    \caption{Stacked $0.2 \sim 2.3$ keV X-ray surface brightness maps for \simba\ (left) and \gadgetx\ (right). The top panels display X-ray emission from warm-hot gas, the middle panels show X-ray emission with halos more massive than $10^{13.5}M_{\odot}/h$ removed, and the bottom panels present X-ray emission from all gas.
    \label{fig:xraymap}}
\end{figure*}

We compute $0.2\sim2.3$ keV X-ray surface brightness maps to facilitate WHIM detection. The maps are centred on the halo centre and stacked. The top panels of \autoref{fig:xraymap} show WHIM-only X-ray maps, which include all WHIM gas regardless of whether it resides within halos. The middle panels display total X-ray emission, similar to the bottom panels, but with halos more massive than $10^{13.5}M_{\odot}$ excluded. The bottom panels present the total X-ray emission from all gas.

Due to the nature of the massive clusters that form our sample, effectively only Coma-like objects are stacked in this analysis. The full region spans $10 \times R_{200c}$. It is evident that \gadgetx\ produces more compact, bright sources, whereas \simba\ exhibits smoother distributions, likely due to stronger feedback mechanisms driving gas particles to larger radii \citep[e.g.][]{Yang2023, Luo2024}. The similarity between the middle and top panels beyond the cluster cores confirms that halo removal effectively isolates WHIM-dominated X-ray emission.

\begin{figure}
    \centering
    \includegraphics[scale=0.4]{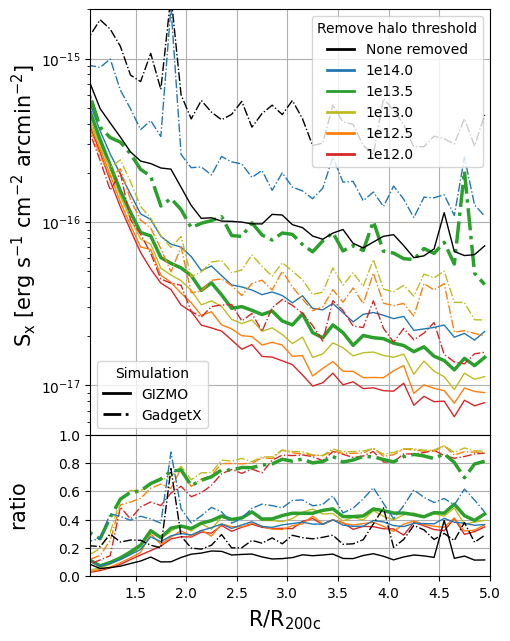}
    \caption{
    The X-ray surface brightness profile obtained via the stacking method (top) and the fractional contribution of the X-ray signal from warm-hot gas (bottom). Line styles represent different simulations, while colours indicate halo removal thresholds. Thickened lines correspond to the removal of halos above $10^{13.5} M_{\odot}/h$.
    \label{fig:xrayprofile}}
\end{figure}

\autoref{fig:xrayprofile} quantifies the radial trends in X-ray emission. After masking pixels containing massive halos, we stack maps following the methodology in \autoref{fig:xraymap}. The average pixel values are then computed in radial bins. Typically, X-ray emission around massive clusters is visually detectable out to $\sim R_{200c}$ or down to a surface brightness of $10^{-15}\,\mathrm{erg\,s^{-1}\,cm^{-2}\,arcmin^{-2}}$, primarily contributed by hot gas \citep[e.g.][]{31Mirakhor2020MNRAS}. Beyond $R_{200c}$, the emission drops below the detection limit of current X-ray telescopes. Notably, differences between the models result in distinct profiles: X-ray emission in \simba\ is significantly lower than in \gadgetx. This discrepancy arises from the density differences shown in \autoref{fig:rhocontour}, where \gadgetx\ maintains higher gas densities in the cluster outskirts, likely due to contributions from low-mass halos, as seen in \autoref{fig:xraymap}. We do not exclude halos with even lower masses, as current X-ray telescopes struggle to detect them beyond the local Universe.

The bottom panel of \autoref{fig:xrayprofile} shows the fractional contribution of the X-ray signal from WHIM gas. Note that the WHIM surface brightness is calculated from all WHIM gas, regardless of their location, following the same approach as the top panels of \autoref{fig:xraymap}. Both simulations exhibit an increase in WHIM contribution when halos above $10^{13.5} M_{\odot}/h$ are removed. As expected, excluding lower-mass halos further reduces the total flux while keeping the WHIM fraction stable. This indicates that a $10^{13.5} M_{\odot}/h$ threshold effectively preserves the signal while isolating WHIM emission. Thus, $10^{13.5} M_{\odot}/h$ appears to be the optimal threshold for detecting WHIM features.

\subsection{The X-ray signal at higher redshift} \label{subsec:obs_redshift}

\begin{figure}
    \centering
    \includegraphics[scale=0.4]{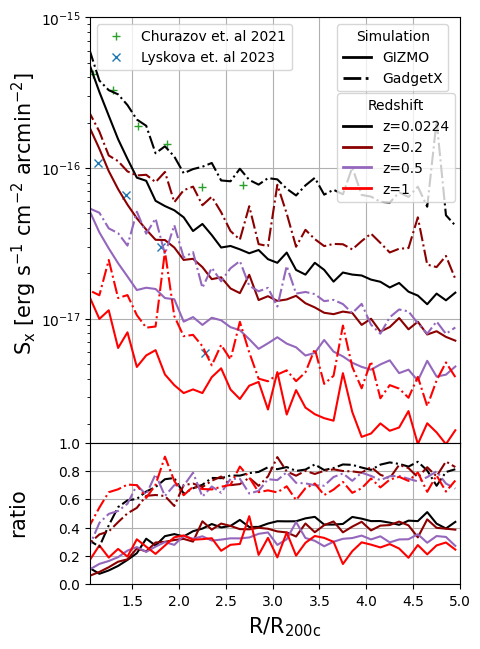}
    \caption{
    Top panel: Stacked X-ray surface brightness profiles for \simba\ and \gadgetx\ at various redshifts.
    Bottom panel: WHIM contribution ratios. Halos above $10^{13.5} M_{\odot}/h$ are removed.
    Green plus symbols (+) represent the Coma cluster surface brightness profile in the $0.4$–$2$ keV band from \cite{27Churazov2021A&A}.
    Blue x-shaped symbols (x) indicate the stacked X-ray profile of clusters at redshifts $0.05$–$0.2$ in the $0.3$–$2.3$ keV band from \cite{Lyskova2023MNRAS}.
    \label{fig:xrayprofile_redshift}}
\end{figure}

\begin{figure}
    \centering
    \includegraphics[scale=0.4]{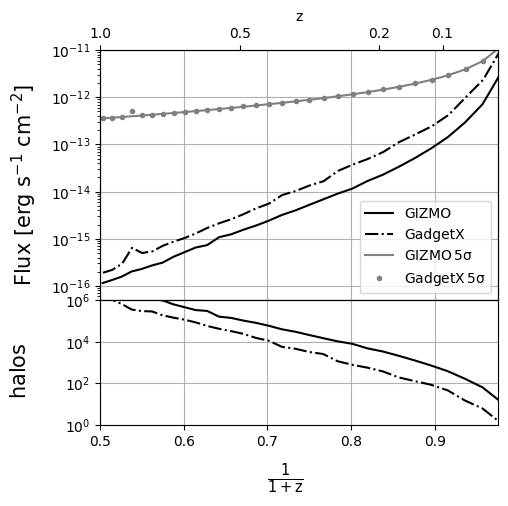}
    \caption{
    The average flux between $2R_{200c}$ and $3R_{200c}$ outside central clusters as a function of redshift, along with their $5\sigma$ significance thresholds (top panel). The bottom panel shows the required number of clusters to achieve a $5\sigma$ detection as a function of expansion factor, $a$. The corresponding redshift is indicated along the top. These results are based on eROSITA observational systematics, assuming an exposure time of 100s.
    \label{fig:xray_redshift}}
\end{figure} 

Increasing redshift correlates with decreasing cluster masses and greater distances. \autoref{fig:xrayprofile_redshift} presents the X-ray profiles at different redshifts. Similarly to \autoref{fig:xrayprofile}, the top panel displays the X-ray surface brightness profiles, while the bottom panel shows the WHIM contribution fractions. Observational data are overplotted: green plus symbols (+) represent the Coma cluster surface brightness profile in the $0.4$–$2.0$ keV band from \cite{27Churazov2021A&A}, and blue x-shaped symbols (x) indicate the stacked X-ray profile of clusters at redshifts $0.05$–$0.2$ in the $0.3$–$2.4$ keV band from \cite{Lyskova2023MNRAS}.

The analysis assumes an energy conversion factor (ECF) of $10^{12} \,\mathrm{cm^2\,erg^{-1}}$ and converts radii using $R_{200c} = 1.6R_{500c}$. Data from these papers have been normalised as per the eROSITA telescope module and are therefore multiplied by 7 to eliminate the effect of normalisation and account for the full array. We find that:

\begin{enumerate}
    \item At all redshifts, the \gadgetx\ model produces a stronger X-ray emission signal than the \simba\ model, even at $z=1$, where \simba\ exhibits significantly higher gas temperatures than \gadgetx. While the \gadgetx\ result at $z \sim 0.02$ aligns well with the Coma profile from \cite{27Churazov2021A&A}, the \simba\ profile at $z \sim 0.2$ closely matches the stacked result from \cite{Lyskova2023MNRAS}, except at $R \sim 2.25\times R_{200c}$. However, we note that our results do not account for any telescope systematics beyond the coarse conversion from $R_{500c}$ to $R_{200c}$.
    \item As expected, X-ray emission decreases significantly with redshift in both simulations, necessitating a large number of clusters for reliable detection at high redshift. 
    \item The WHIM contribution remains largely independent of redshift (bottom panel), despite a factor of two difference between \gadgetx\ and \simba. This consistency simplifies the interpretation of the WHIM fraction derived from X-ray emission.
    \item Radial profiles flatten beyond $2\times R_{200c}$, which is consistent with previous studies. Therefore, we analyse the X-ray flux between $2R_{200c}$ and $3R_{200c}$, a range that approximately corresponds to the splashback radius and the shock radius \citep{Zhangming2025}, for WHIM signal characterization in \autoref{subsec:obs_detect}.
\end{enumerate}

\subsection{Detectability} \label{subsec:obs_detect}

For observational feasibility, it is crucial to determine the number of clusters that must be stacked to achieve a desired X-ray signal-to-noise ratio. Using data from the first eROSITA All-Sky Survey \citep[eRASS1;][]{Merloni2024} and the eROSITA Final Equatorial-Depth Survey \citep[eFEDS;][]{Brunner2022, Liu2022}, we estimate the background level of eROSITA surveys to be $BG = 0.0059\, \mathrm{counts\, s^{-1} \, arcmin^{-2}}$. Assuming an exposure time of $t=100$~s and using simulated cluster angular sizes, we compute the theoretical $5\sigma$ detection threshold, as shown in \autoref{fig:xray_redshift}. By comparing the mean simulated X-ray fluxes, the required number of clusters to achieve $5\sigma$ significance is determined by solving the equation:
\begin{equation} \rm SNR = \frac{Signal}{\sqrt{Signal+Noise}} \label{eq_stackhalos}\,, \end{equation}
where
\begin{equation} \rm Signal = \sum F \times ECF \times time \label{eq_signal}\,, \end{equation}
and
\begin{equation} \rm Noise = \sum BG \times time \times A_c \label{eq_noise}\,. \end{equation}
Here, $F$ is the average flux in $\mathrm{erg\,s^{-1}\,cm^{-2}}$, $ECF$ is the energy conversion factor, taken as $1.074\times10^{12}\,\mathrm{cm^2\,erg^{-1}}$ for the $0.2\sim2.3$ keV band \citep{Brunner2022}, and $A_c$ is the solid angle of the statistical area.

As shown in \autoref{fig:xray_redshift}, the average flux from both models decreases significantly with redshift, dropping from approximately $3 (8) \times 10^{-12} \,\mathrm{erg\,s^{-1}\,cm^{-2}}$ for \simba\ (\gadgetx) to around $10^{-16} \,\mathrm{erg\,s^{-1}\,cm^{-2}}$. Consequently, the number of clusters required to achieve a $5\sigma$ detection is expected to increase exponentially. Since \simba\ produces a lower X-ray flux due to its lower gas density, it is unsurprising that a larger number of clusters must be stacked to reach $5\sigma$ significance.

Using the expectations from \autoref{fig:xray_redshift}, we predict the achievable signal-to-noise ratio (SNR) based on the eRASS1 galaxy cluster catalogue \citep{Bulbul2024A&A}. Our sample selection criteria includes clusters with $0.1 < z < 0.5$, $L_{ext} > 6$, and $P_{cont} < 0.3$. Cluster masses are converted using the approximation $M_{200c} \sim 3M_{500c}$, and only clusters with $M_{200c} > 10^{14.5} M_{\odot}$ are retained to match our simulated cluster sizes. Signal and noise values are interpolated from the simulated results based on cluster redshifts, and exposure times are taken directly from the catalogue. After applying these criteria, 3802 clusters remain for stacking. We estimate an SNR of $\sim 7.07$ for the \simba\ model and $\sim 20.81$ for the \gadgetx\ model. These values would approximately double if eRASS-4 data were used, providing an even more optimised detection of the WHIM.

\section{Conclusions} \label{sec:con}

Using \theth\ simulations, we investigate detection methods for missing baryons, primarily in the form of WHIM, in cluster outskirts. By applying a stacking methodology, we generate averaged X-ray surface brightness maps and profiles and estimate the required sample sizes for WHIM detection with eRASS1. Both the \simba\ and \gadgetx\ models are analysed to assess differences between simulation models and to ensure the robustness of our results. The main findings are summarised as follows:

\begin{itemize}
    \item 
    Warm-hot gas serves as the primary reservoir of the missing baryons due to its intermediate temperatures ($10^5 \sim 10^7$~K) and low densities. It becomes the dominant component beyond $2R_{200c}$ in cluster outskirts. Both \simba\ and \gadgetx\ models predict similar WHIM fractions, but \gadgetx\ exhibits higher gas densities, whereas \simba\ tends to have higher temperatures in the cluster outskirts.
    \item 
    At lower redshift, the density of warm-hot gas is lower because of the expansion of the universe. And the fraction of the warm-hot element is also lower because of the evolution of the structures, which may heat the gas within $\sim 2\times R_{200c}$, while the WHIM mass fraction is more constant beyond that radius.
    \item 
    At lower redshifts, the density of warm-hot gas decreases due to cosmic expansion. The fraction of warm-hot gas is also lower, as structure formation heats the gas within $\sim 2\times R_{200c}$. Beyond this radius, however, the WHIM mass fraction remains relatively constant.
    \item 
    The warm-hot gas fraction is highest between $2R_{200c}$ and $3R_{200c}$. In this region, X-ray fluxes for \gadgetx\ clusters range from $2\times10^{-16} \,\mathrm{erg\,s^{-1}\,cm^{-2}}$ at $z=1$ to $10^{-11} \,\mathrm{erg\,s^{-1}\,cm^{-2}}$ at $z=0.0224$. \simba\ fluxes are approximately 50\% lower than those of \gadgetx.
    \item 
    Using the eRASS1 cluster catalogue and data, we predict a $\mathrm{SNR} \sim 7.07$ for the \simba\ model and $\mathrm{SNR} \sim 20.81$ for the \gadgetx\ model.
\end{itemize}

Finally, we note that our X-ray emission calculations are theoretical and do not account for observational corrections such as instrument response.
These factors are closely tied to specific observational setups, and we leave their incorporation to future work focused on detecting warm-hot gas using targeted observational surveys.

\begin{acknowledgements}
We thank Elena Rasia, Sara Santoni, Yunliang Zheng and Xiaoyuan Zhang for useful discussions.

This work is supported by the National Natural Science Foundation of China Nos. 12192224.
RJL is supported by the China Scholarship Council.
WC is supported by Atracci\'{o}n de Talento Contract no. 2020-T1/TIC19882 granted by the Comunidad de Madrid in Spain, and the science research grants from the China Manned Space Project. He also thanks the Ministerio de Ciencia e Innovaci\'{o}n (Spain) for financial support under Project grant PID2021-122603NB-C21 and ERC: HORIZON-TMA-MSCA-SE for supporting the LACEGAL-III Latin American Chinese European Galaxy Formation Network) project with grant number 101086388.
HYW acknowledgements supports from CAS Project for Young Scientists in Basic Research, Grant No. YSBR-062, the science research grants from the China Manned Space Project with CMS-CSST-2021-A03 and Cyrus Chun Ying Tang Foundations.
AS is also supported by Atracci\'{o}n de Talento Contract no. 2020-T1/TIC19882 granted by the Comunidad de Madrid in Spain.

The authors acknowledge The Red Española de Supercomputaci\'on for granting computing time for running the hydrodynamical and DMO simulations of {\sc The Three Hundred} galaxy cluster project in the Marenostrum supercomputer at the Barcelona Supercomputing Center and Cibeles Supercomputers through various of RES grants. 

For the purpose of open access, the author has applied a Creative Commons Attribution (CC BY) licence to any Author Accepted Manuscript version arising from this submission.

\end{acknowledgements}
\bibliographystyle{aa} 
\bibliography{main} 

\end{document}